\documentclass[preprintnumbers,amsmath,amssymbm,prd]{revtex4}
\usepackage{epsfig}
\usepackage{graphicx}

\begin{document}
\title{The Hawking cascades of gravitons from higher-dimensional Schwarzschild black
holes}
\author{Shahar Hod}
\affiliation{The Ruppin Academic Center, Emeq Hefer 40250, Israel}
\affiliation{ } \affiliation{The Hadassah Institute, Jerusalem
91010, Israel}
\date{\today}

\begin{abstract}
\ \ \ It has recently been shown that the Hawking evaporation
process of $(3+1)$-dimensional Schwarzschild black holes is
characterized by the dimensionless ratio
$\eta\equiv\tau_{\text{gap}}/\tau_{\text{emission}}\gg1$, where
$\tau_{\text{gap}}$ is the characteristic time gap between the
emissions of successive Hawking quanta and $\tau_{\text{emission}}$
is the characteristic timescale required for an individual Hawking
quantum to be emitted from the Schwarzschild black hole. This strong
inequality implies that the Hawking cascade of gravitons from a
$(3+1)$-dimensional Schwarzschild black hole is extremely {\it
sparse}. In the present paper we explore the semi-classical Hawking
evaporation rates of {\it higher}-dimensional Schwarzschild black
holes. We find that the dimensionless ratio
$\eta(D)\equiv{{\tau_{\text{gap}}}/{\tau_{\text{emission}}}}$, which
characterizes the Hawking emission of gravitons from the
$(D+1)$-dimensional Schwarzschild black holes, is a {\it decreasing}
function of the spacetime dimension. In particular, we show that
higher-dimensional Schwarzschild black holes with $D\gtrsim 10$ are
characterized by the relation $\eta(D)<1$. Our results thus imply
that, contrary to the $(3+1)$-dimensional case, the characteristic
Hawking cascades of gravitons from these higher-dimensional black
holes have a {\it continuous} character.
\end{abstract}
\bigskip
\maketitle

\section{Introduction}

Analyzing the dynamics of quantum fields in curved black-hole
spacetimes, Hawking \cite{Haw} has reached the intriguing conclusion
that black holes have a well defined temperature \cite{NoteBek,Bek}.
In particular, Hawking \cite{Haw} has revealed that semi-classical
black holes are characterized by quantum emission spectra which have
distinct thermal features \cite{Notegrey}. This remarkable
prediction is certainly one of the most important outcomes of the
interplay between quantum field theory and classical general
relativity.

It has recently been shown \cite{Vis} that the semi-classical
Hawking radiation flux out of a $(3+1)$-dimensional Schwarzschild
black hole is extremely sparse [see also \cite{BekMuk,Mak,Hod1} for
earlier discussions of this characteristic property of the
$(3+1)$-dimensional Schwarzschild black hole]. In particular, the
Hawking emission of gravitons out of a $(3+1)$-dimensional
Schwarzschild black hole is characterized by the dimensionless large
ratio \cite{Vis}
\begin{equation}\label{Eq1}
\eta\equiv{{\tau_{\text{gap}}}\over{\tau_{\text{emission}}}}\gg1\ ,
\end{equation}
where $\tau_{\text{gap}}$ is the characteristic time gap between the
emissions of successive black-hole gravitational quanta [see Eqs.
(\ref{Eq8}) and (\ref{Eq9}) below], and $\tau_{\text{emission}}$ is
the characteristic timescale required for an individual Hawking
quantum to be emitted from the black hole [see Eq. (\ref{Eq10})
below].

The dimensionless large ratio (\ref{Eq1}) implies that the
semi-classical Hawking evaporation of $(3+1)$-dimensional
Schwarzschild black holes is indeed sparse. That is, the
characteristic time gap between the emissions of successive
gravitational quanta out of an evaporating $(3+1)$-dimensional
Schwarzschild black hole is very large on the natural timescale
$2\pi/\omega$ [see Eq. (\ref{Eq10}) below] set by the characteristic
energy (frequency) of the emitted Hawking quanta. The characteristic
large ratio (\ref{Eq1}) therefore suggests a simple physical picture
in which an evaporating $(3+1)$-dimensional Schwarzschild black hole
\cite{NoteHK,Hodkerr} typically emits Hawking quanta one at a time
\cite{Vis,BekMuk,Mak,Hod1}.

One naturally wonders whether the strong inequality (\ref{Eq1}),
which characterizes the Hawking evaporation process of the
$(3+1)$-dimensional Schwarzschild black hole, is a generic feature
of all $(D+1)$-dimensional Schwarzschild black-hole spacetimes? In
order to answer this interesting question, we shall explore in this
paper the semi-classical Hawking evaporation of higher-dimensional
Schwarzschild black holes. Below we shall show that the
dimensionless ratio
$\eta(D)\equiv{{\tau_{\text{gap}}}/{\tau_{\text{emission}}}}$, which
characterizes the Hawking evaporation process of $(D+1)$-dimensional
Schwarzschild black holes, is a {\it decreasing} function of the
spacetime dimension. In particular, we shall show that
higher-dimensional Schwarzschild black holes with $D\gtrsim 10$ are
characterized by the relation $\eta(D)<1$. Thus, our analysis (to be
presented below) reveals the fact that the Hawking cascades of
gravitons from these higher-dimensional evaporating black holes have
a {\it continuous} character.

\section{The Hawking evaporation process of $(D+1)$-dimensional Schwarzschild black
holes}

We study the semi-classical Hawking emission of gravitational quanta
by higher-dimensional Schwarzschild black holes. The characteristic
Bekenstein-Hawking temperature of an evaporating $(D+1)$-dimensional
Schwarzschild black hole is given by \cite{Noteunit}
\begin{equation}\label{Eq2}
T^{D}_{\text{BH}}={{(D-2)\hbar}\over{4\pi r_{\text{H}}}}\  ,
\end{equation}
where $r_{\text{H}}$ is the black-hole horizon radius
\cite{Noterh,SchTang,Notekk,Ober}.

The radiation flux (that is, the number of quanta emitted per unit
of time) and the radiation power (that is, the energy emitted per
unit of time) for one bosonic degree of freedom out of a
$(D+1)$-dimensional Schwarzschild black hole are given respectively
by the integral relations \cite{Haw,Page,ZuKa}
\begin{equation}\label{Eq3}
{\cal
F}^{D}_{\text{BH}}={{1}\over{(2\pi)^D}}\sum_{j}{\int_0^{\infty}}dV_D(\omega)
{{\Gamma}\over{e^{\hbar\omega/T_{\text{BH}}}-1}}\
\end{equation}
and
\begin{equation}\label{Eq4}
{\cal
P}^{D}_{\text{BH}}={{1}\over{(2\pi)^D}}\sum_{j}{\int_0^{\infty}}dV_D(\omega)
{{\Gamma\hbar\omega}\over{e^{\hbar\omega/T_{\text{BH}}}-1}}\ ,
\end{equation}
where $j$ denotes the angular harmonic indices of the emitted field
mode. The frequency-dependent coefficients
$\Gamma=\Gamma(\omega;j,D)$ are the dimensionless greybody factors
\cite{Page} which quantify the partial scattering of the emitted
field modes by the effective curvature potential that surrounds the
$(D+1)$-dimensional Schwarzschild black hole. These factors are
determined by the higher-dimensional version of the Regge-Wheeler
equation \cite{CKK}
\begin{equation}\label{EqN1}
\Big({{d^2}\over{dr^2_*}}+\omega^2-V\Big)\phi=0\  ,
\end{equation}
where $r_*$ is a `tortoise' radial coordinate which is determined by
the relation $dr_*/dr=[1-(r_{\text{H}}/r)^{D-3}]^{-1}$. The
effective curvature potential in the Schr\"odinger-like wave
equation (\ref{EqN1}) is given by \cite{CKK}
\begin{equation}\label{EqN2}
V(r;D)=\Big[1-\Big({{r_{\text{H}}}\over{r}}\Big)^{D-3}\Big]\Big[{{l(l+D-2)+(D-1)(D-3)/4}\over{r^2}}
+{{(1-p^2)(D-1)^2r^{D-2}_{\text{H}}}\over{4r^D}}\Big]\ ,
\end{equation}
where $l$ is the angular harmonic index of the perturbation mode and
$p=0,2$ for gravitational tensor perturbations and gravitational
vector perturbations, respectively \cite{CKK,Notecom}.

Substituting into (\ref{Eq3}) and (\ref{Eq4}) the expression
\begin{equation}\label{Eq5}
dV_D(\omega)=[2\pi^{D/2}/\Gamma(D/2)]\omega^{D-1}d\omega\
\end{equation}
for the D-dimensional volume in frequency-space of the shell
$(\omega,\omega+d\omega)$, one finds
\begin{equation}\label{Eq6}
{\cal
F}^D_{\text{BH}}={{1}\over{2^{D-1}\pi^{D/2}\Gamma(D/2)}}\sum_{j}{\int_0^{\infty}}\Gamma
{{\omega^{D-1}}\over{{e^{\hbar\omega/T_{\text{BH}}}-1}}} d\omega\
\end{equation}
and
\begin{equation}\label{Eq7}
{\cal
P}^D_{\text{BH}}={{\hbar}\over{2^{D-1}\pi^{D/2}\Gamma(D/2)}}\sum_{j}{\int_0^{\infty}}\Gamma
{{\omega^D}\over{{e^{\hbar\omega/T_{\text{BH}}}-1}}} d\omega\
\end{equation}
for the semi-classical Hawking radiation flux and the semi-classical
Hawking radiation power which characterize the evaporating
$(D+1)$-dimensional Schwarzschild black holes.

\section{The characteristic timescales of the Hawking evaporation
process}

An important timescale which characterizes the Hawking evaporation
process of the $(D+1)$-dimensional Schwarzschild black holes is
given by the time gap between the emissions of successive Hawking
quanta. There are several distinct (though closely related) ways to
quantify this fundamental time scale. Here we shall use two natural
definitions for this characteristic time gap:
\newline
(1) One can use the reciprocal of the black-hole radiation flux
(\ref{Eq6}) in order to quantify the characteristic time gap between
the emissions of successive Hawking quanta. That is,
\begin{equation}\label{Eq8}
\tau^{(1)}_{\text{gap}}={{1}\over{{\cal F}^D_{\text{BH}}}}\  .
\end{equation}
\newline
(2) One can also use the reciprocal of the black-hole radiation
power (\ref{Eq7}) and the characteristic peak frequency
$\omega_{\text{peak}}$ of the semi-classical Hawking radiation
spectrum in order to quantify the characteristic time gap between
the emissions of successive Hawking quanta. That is,
\begin{equation}\label{Eq9}
\tau^{(2)}_{\text{gap}}={{\omega_{\text{peak}}}\over{{\cal
P}^D_{\text{BH}}}}\ .
\end{equation}
Below we shall show that the characteristic time gaps obtained from
these two definitions [Eqs. (\ref{Eq8}) and (\ref{Eq9})] are of the
same order of magnitude.

A distinct timescale which characterizes the Hawking evaporation
process of the $(D+1)$-dimensional Schwarzschild black holes is
given by the time $\tau_{\text{emission}}$ required for an
individual Hawking quantum to be emitted from the evaporating black
hole. This fundamental timescale can be bounded from below by the
time-period it takes to the characteristic wave field emitted from
the black hole to complete a full oscillation cycle
\cite{Vis,Hodkerr}. That is,
\begin{equation}\label{Eq10}
\tau_{\text{emission}}\geq\tau_{\text{oscillation}}={{2\pi}\over{\omega_{\text{peak}}}}\
.
\end{equation}

Using the natural timescales (\ref{Eq8}), (\ref{Eq9}), and
(\ref{Eq10}), one can define the fundamental dimensionless ratio
\begin{equation}\label{Eq11}
\eta^{(i)}\equiv{{\tau^{(i)}_{\text{gap}}}\over{\tau_{\text{emission}}}}\
\end{equation}
which provides important information about the Hawking evaporation
process of the semi-classical black holes \cite{Notepk}. In
particular, physical situations which are characterized by the
relation $\eta\gg1$ describe Hawking evaporation processes which are
extremely {\it sparse} (that is, the individual Hawking quanta
emitted from the black hole are well separated in time), whereas
physical situations which are characterized by the relation
$\eta\ll1$ describe Hawking evaporation processes which are
effectively {\it continuous}.

In the next sections we shall investigate the functional dependence
of the dimensionless ratios
$\eta^{(i)}(D)\equiv{{\tau^{(i)}_{\text{gap}}}/{\tau_{\text{emission}}}}$
on the spacetime dimension $D+1$ of the evaporating black-hole
spacetime.

\section{The $(3+1)$-dimensional case}

For the emission of gravitational Hawking quanta from a
$(3+1)$-dimensional Schwarzschild black hole \cite{Page} one finds
the characteristic dimensionless ratios
\begin{equation}\label{Eq12}
\eta^{(1)}(D=3)=5175.8\gg1\ \ \ \ \ \text{and}\ \ \ \ \
\eta^{(2)}(D=3)=5371.5\gg1\ \ .
\end{equation}
As emphasized earlier, these remarkably large ratios imply that the
Hawking cascade from the evaporating $(3+1)$-dimensional
Schwarzschild black hole is extremely {\it sparse}
\cite{Vis,BekMuk,Mak,Hod1}. In other words, the dimensionless large
ratios (\ref{Eq12}) imply that, on average, an evaporating
$(3+1)$-dimensional Schwarzschild black hole emits gravitational
quanta which are well separated in time \cite{Vis,BekMuk,Mak,Hod1}.

\section{Higher-dimensional black holes: Intermediate $D$-values}

In the previous section we have seen that the $(3+1)$-dimensional
Schwarzschild black hole is characterized by large values of the
dimensionless ratios $\eta^{(i)}(D=3)$ [see Eq. (\ref{Eq12})]. In
the present section we shall show that the dimensionless ratios
$\eta^{(i)}(D)$, which characterize the Hawking evaporation process
of $(D+1)$-dimensional Schwarzschild black holes, are {\it
decreasing} functions of the spacetime dimension.

The semi-classical Hawking emission of gravitons from
higher-dimensional Schwarzschild black holes was investigated
numerically in \cite{CKK}. In Table \ref{Table1} we display the
numerically computed dimensionless ratios
${{\tau^{(i)}_{\text{gap}}}/{\tau_{\text{oscillation}}}}$ which
characterize the Hawking emission of gravitons from
higher-dimensional Schwarzschild black holes with intermediate
$D$-values \cite{Notebo}. One finds from Table \ref{Table1} that the
characteristic dimensionless ratios
${{\tau^{(i)}_{\text{gap}}}/{\tau_{\text{oscillation}}}}$ of the
$(D+1)$-dimensional Schwarzschild black holes are {\it decreasing}
functions of the spacetime dimension $D+1$.

In particular, we find that Schwarzschild black holes in the regime
$D=O(10)$ are characterized by the dimensionless ratios
${{\tau^{(i)}_{\text{gap}}}/{\tau_{\text{oscillation}}}}=O(1)$.
These higher-dimensional black holes therefore mark the boundary
between sparse (that is, with $\eta^{(i)}\gg1$) Hawking cascades of
gravitons and continuous (that is, with $\eta^{(i)}\ll1$) Hawking
cascades of gravitons.


\begin{table}[htbp]
\centering
\begin{tabular}{|c|c|c|c|}
\hline $D+1$ & \ \ 5 \ \ & \ \ 8 \ \ & \ \ 11 \ \ \ \\
\hline \ \ ${{\tau^{(1)}_{\text{gap}}}/{\tau_{\text{oscillation}}}}$
\ \ &\ \ \ 120.9\ \ \ \ &\ \ \ 3.42\ \ \ \ &\ \ \ 0.57\ \ \ \ \\
\hline \ \ ${{\tau^{(2)}_{\text{gap}}}/{\tau_{\text{oscillation}}}}$
\ \ &\ \ \ 105.1\ \ \ \ &\ \ \ 4.22\ \ \ \ &\ \ \ 0.56\ \ \ \ \\
\hline
\end{tabular}
\caption{The dimensionless ratios
${{\tau^{(i)}_{\text{gap}}}/{\tau_{\text{oscillation}}}}$ which
characterize the semi-classical Hawking emission of gravitons from
$(D+1)$-dimensional Schwarzschild black holes \cite{CKK}. Here
$\tau^{(i)}_{\text{gap}}$ is the characteristic time gap between the
emissions of successive Hawking quanta from the higher-dimensional
black hole [see Eqs. (\ref{Eq8}) and (\ref{Eq9})] and
$\tau_{\text{oscillation}}$ is the characteristic oscillation period
of the emitted wave field [see Eq. (\ref{Eq10})]. One finds that the
characteristic dimensionless ratios
${{\tau^{(i)}_{\text{gap}}}/{\tau_{\text{oscillation}}}}$ are {\it
decreasing} functions of the spacetime dimension $D+1$.
} \label{Table1}
\end{table}

\section{Higher-dimensional black holes: The large-$D$ regime}

The D-dimensional frequency distribution
$\omega^{D}/(e^{\hbar\omega/T_{\text{BH}}}-1)$ [see Eq. (\ref{Eq7})]
is characterized by the peak frequency
\begin{equation}\label{Eq13}
{{\hbar\omega_{\text{peak}}}\over{T^{D}_{\text{BH}}}}=D+W(-De^{-D})\
,
\end{equation}
where $W(x)$ is the Lambert function. Using the small argument
relation $W(x\to0)\to0$ and taking cognizance of the
Bekenstein-Hawking temperature (\ref{Eq2}), one finds the
characteristic strong inequality \cite{Hoddd}
\begin{equation}\label{Eq14}
\omega_{\text{peak}}\times
r_{\text{H}}={{D^2}\over{4\pi}}[1+O(D^{-1})]\gg1\
\end{equation}
in the large $D\gg1$ regime.

The strong inequality (\ref{Eq14}) reflects the fact that, for
$(D+1)$-dimensional Schwarzschild black holes in the large $D\gg1$
regime, the characteristic wavelengths in the semi-classical Hawking
emission spectrum are very {\it short} as compared to the curvature
radius of the corresponding black-hole spacetime. As demonstrated in
\cite{Hoddd}, this fact implies that the semi-classical Hawking
evaporation of these higher-dimensional black holes is described
extremely well by the eikonal (geometric-optics) approximation
\cite{Notego}. In particular, one finds \cite{Hoddd}
\begin{equation}\label{Eq15}
{\cal F}^{D}_{\text{BH}}\times
r_{\text{H}}={{(D+1)(D-2)}\over{2}}\Big({{D-2}\over{4\pi}}\Big)^{D}\Big({{D}\over{2}}\Big)^{{D-1}\over{D-2}}
\Big({{D}\over{D-2}}\Big)^{{D-1}\over{2}}{{\zeta(D)}\over{\pi}}
\end{equation}
for the semi-classical Hawking radiation flux, and \cite{Hoddd}
\begin{equation}\label{Eq16}
{\cal P}^{D}_{\text{BH}}\times
r^2_{\text{H}}={{(D+1)(D-2)}\over{2}}\Big({{D-2}\over{4\pi}}\Big)^{D+1}\Big({{D}\over{2}}\Big)^{{D-1}\over{D-2}}
\Big({{D}\over{D-2}}\Big)^{{D-1}\over{2}}{{D\zeta(D+1)\hbar}\over{\pi}}
\end{equation}
for the semi-classical Hawking radiation power in the large-D regime
[which, as explained above, corresponds to the {\it short}
wavelengths (geometric-optics) approximation, see Eq. (\ref{Eq14})].
Taking the asymptotic large $D\gg1$ limit in Eqs. (\ref{Eq15}) and
(\ref{Eq16}), one obtains the compact asymptotic expressions
\begin{equation}\label{Eq15N}
{\cal F}^{D}_{\text{BH}}\times
r_{\text{H}}={{(4\pi)^2}\over{e}}\Big({{D}\over{4\pi}}\Big)^{D+3}\ \
\ \text{for}\ \ \ D\gg1\
\end{equation}
and
\begin{equation}\label{Eq16N}
{\cal P}^{D}_{\text{BH}}\times
r^2_{\text{H}}=\hbar{{(4\pi)^3}\over{e}}\Big({{D}\over{4\pi}}\Big)^{D+5}\
\ \ \text{for}\ \ \ D\gg1\  .
\end{equation}

Taking cognizance of Eqs. (\ref{Eq8}), (\ref{Eq9}), (\ref{Eq10}),
(\ref{Eq15N}), and (\ref{Eq16N}), one finds the remarkably {\it
small} dimensionless ratios
\begin{equation}\label{Eq17}
\eta^{(1)}(D\gg1)=\eta^{(2)}(D\gg1)={{e}\over{8\pi^2}}\Big({{4\pi}\over{D}}\Big)^{D+1}\ll1
\end{equation}
which characterize the evaporating $(D+1)$-dimensional Schwarzschild
black holes in the large-D limit \cite{Notemat}. These relations
(and, in particular, the strong inequalities
$\tau^{(i)}_{\text{gap}}\ll\tau_{\text{emission}}$) imply that the
characteristic Hawking cascades from these higher-dimensional black
holes have a {\it continuous} character.

\section{Summary}

It has long been known \cite{Vis,BekMuk,Mak,Hod1} that an
evaporating $(3+1)$-dimensional Schwarzschild black hole is
characterized by an extremely {\it sparse} Hawking radiation flux.
In particular, the dimensionless large ratios
$\eta^{(i)}\equiv\tau^{(i)}_{\text{gap}}/\tau_{\text{emission}}=O(10^3)$
which characterize the $(3+1)$-dimensional Schwarzschild black hole
[see Eq. (\ref{Eq12})] imply a simple physical picture in which the
evaporating black hole typically emits Hawking quanta one at a time
\cite{Vis,BekMuk,Mak,Hod1}.

In the present paper we have analyzed the Hawking emission rates of
higher-dimensional Schwarzschild black holes. It was shown that the
dimensionless ratios
$\eta^{(i)}(D)\equiv{{\tau^{(i)}_{\text{gap}}}/{\tau_{\text{emission}}}}$,
which characterize the semi-classical Hawking emission of gravitons
from the $(D+1)$-dimensional Schwarzschild black holes, are {\it
decreasing} functions of the spacetime dimension. In particular, we
have shown that higher-dimensional Schwarzschild black holes with
$D\gtrsim 10$ are characterized by the relation $\eta^{(i)}(D)<1$.
This fact implies that the corresponding Hawking cascades from these
higher-dimensional black holes have a continuous character with the
property $\tau_{\text{gap}}<\tau_{\text{emission}}$.

Moreover, we have shown that the semi-classical Hawking emission
spectra of higher-dimensional Schwarzschild black holes in the
large-D regime are characterized by the strong inequality [see Eq.
(\ref{Eq17})]
\begin{equation}\label{Eq18}
\tau_{\text{gap}}(D\gg1)\ \ll\ \tau_{\text{emission}}(D\gg1)\  .
\end{equation}
Our results therefore imply that, contrary to the
$(3+1)$-dimensional case, the characteristic Hawking cascades from
these higher-dimensional black holes \cite{Notetn} have a {\it
continuous} character.

\bigskip
\noindent {\bf ACKNOWLEDGMENTS}
\bigskip

This research is supported by the Carmel Science Foundation. I thank
Yael Oren, Arbel M. Ongo, Ayelet B. Lata, and Alona B. Tea for
stimulating discussions.


\begin{thebibliography}{99}
\bibitem{NoteBek} It is worth emphasizing that this black-hole temperature
is closely related to the black-hole entropy predicted few years
earlier by Bekenstein \cite{Bek}.
\bibitem{Notegrey} It is worth emphasizing that the Hawking black-hole emission spectra of
evaporating semi-classical black holes are not exactly thermal. In
particular, the effective curvature potential which surrounds an
evaporating black hole [see Eqs. (\ref{Eq3}) and (\ref{Eq4}) below]
blocks the low-energy part of the black-hole radiation spectrum.
\bibitem{Noteunit} We use gravitational units in which $G=c=k_{\text{B}}=1$.
\bibitem{Noterh} The horizon-radius of a $(D+1)$-dimensional Schwarzschild black hole of mass $M$ is
given by \cite{SchTang} $r_{\text{H}}={[{{16\pi
M}/{(D-1)A_{D-1}}}]}^{1/{(D-2)}}$, where
$A_{D-1}={{2\pi^{D/2}}/{\Gamma(D/2)}}$ is the generalized area of a
unit $(D-1)$-sphere.
\bibitem{Notekk} It is worth noting that the spherically symmetric Tangherlini black
holes \cite{SchTang} serve here as toy models for black-hole
solutions in higher dimensional spacetimes. It should be emphasized,
however, that in higher-dimensional Kaluza-Klein models one finds
many phases of black-hole solutions. These higher-dimensional
solutions include non-uniform black strings, localized black holes,
bubble-black hole sequences, and multi-black hole configurations,
see \cite{Ober} for an excellent review and a detailed list of
references.
\bibitem{Notecom} The effective curvature potential in (\ref{EqN1}) for gravitational scalar
perturbations has a rather cumbersome form which is given in
\cite{CKK}.
\bibitem{Notepk} Using in equation (\ref{Eq11}) the average frequency $\omega_{\text{ave}}$ of the
Hawking black-hole radiation spectrum instead of the characteristic
peak frequency $\omega_{\text{peak}}$ of the Hawking black-hole
radiation spectrum would merely change the values of the
dimensionless ratios $\eta^{(i)}$ by a factor of order unity.
\bibitem{Notebo} Note that the numerical data presented in Table \ref{Table1}
provide upper bounds on the values of the characteristic
dimensionless ratios
$\eta^{(i)}(D)\equiv{{\tau^{(i)}_{\text{gap}}}/{\tau_{\text{emission}}}}$
[see Eq. (\ref{Eq10})].
\bibitem{Notego} Note, in particular, that the effective curvature potential (\ref{EqN2})
is characterized by the asymptotic large $D\gg1$ behavior
$V_{\text{max}}(D\gg1)=O(D^2/r^2_{\text{H}})$. One therefore finds
that the Hawking radiation spectrum is characterized by the strong
inequality $\omega^2_{\text{peak}}\gg V_{\text{max}}$ [see Eq.
(\ref{Eq14})] in the large $D\gg1$ regime. This characteristic
large-D relation implies that emitted Hawking quanta in the regime
$l\ll D$ are almost unaffected by the effective scattering potential
(\ref{EqN2}). This fact supports the use of the geometric-optics
(short wavelengths) approximation in the asymptotic large-D regime.
\bibitem{Notemat} It is worth noting that the large-D
geometric-optics relation (\ref{Eq17}) provides the correct order of
magnitude for the dimensionless ratios $\eta^{(1)}$ and $\eta^{(2)}$
already for $D=10$. Specifically, one finds from (\ref{Eq17})
$\eta^{(1)}(D=10)=\eta^{(2)}(D=10)\simeq 0.42$, which is a factor
$\sim 1.3$ smaller than the numerically computed values for the case
$D=10$, see Table \ref{Table1}.
\end{thebibliography}
\end{document}